\documentclass[footinbib,twocolumn,showpacs,amsmath,amstex,amssymb,mathfonts,superscriptaddress,prl]{revtex4}
\usepackage{graphicx}
\usepackage{color}
\usepackage{bm}

\usepackage{graphicx}
\usepackage{color}
\usepackage{bm}
\usepackage{amsmath}
\usepackage{amssymb}
\usepackage{amsthm}
\usepackage{amsfonts}
\usepackage{multirow}
\usepackage{makecell}
\usepackage{float}

\usepackage{bbm}

\begin{document}

\title{Aspects of Three-body Interactions in Generic Fractional Quantum Hall Systems and Impact of Galilean Invariance Breaking}
\author{Bo Yang} 
\affiliation{Division of Physics and Applied Physics, Nanyang Technological University, Singapore 637371.}
\affiliation{Institute of High Performance Computing, A*STAR, Singapore, 138632.}

\date{\today}
\begin{abstract}
We derive full analytic expressions of three-body interactions from Landau level (LL) mixing in fractional quantum Hall (FQH) systems with Schrieffer-Wolff transformation. The formalism can be applied to any LL, and to very general systems without rotational or Galilean invariance. We illustrate how three-body pseudopotentials (PPs) can be readily computed from the analytical expressions for a wide variety of different systems, and show that for realistic systems, softening the bare Coulomb interactions (e.g. finite thickness or screening) can significantly suppress three-body interactions. More interestingly, for experimental systems without Galilean invariance (which is common for real materials), there is strong evidence that higher orders in band dispersion can drive the Moore-Read state from anti-Pfaffian to Pfaffian phase. Our analysis points to the importance of the realistic band structure details to the non-Abelian topological phases, and the analytical expressions we derived can also be very useful for high fidelity numerical computations.
\end{abstract}

\maketitle 

The fractional quantum Hall (FQH) system is a class of two-dimensional strongly correlated electron systems in which the kinetic energy is almost completely quenched by strong magnetic field and low temperature\cite{tsui,prange}. A prototypical example of strongly correlated topological phases, FQH states are the breeding grounds for exotic and potentially highly useful physical phenomena, especially with non-Abelian quasiparticle excitations\cite{dassarma,halperin,shtengel,nayak}. In the FQH regime, electrons partially fill the topmost LL and stay there most of the time due to the large energy gaps between LLs. More interestingly, while fundamentally only two-body Coulomb interaction can exist between electrons, there can be effective three or more body interactions due to the quenching of the kinetic energies and virtual excitations into higher LLs\cite{nayak2}. Such effective many-body interactions are of utmost importance to some of the most interesting, non-Abelian topological phases in FQH systems\cite{MR, Wilczek91, RR}, with rich physical implications and potential practical applications including the design of topological quantum computers.

A number of detailed perturbative and numerical calculations for the effects of LL mixing from realistic interactions have been reported in literature\cite{nayak2, rezayi,simon,zaletel,jain,troyer,nayak3,macdonald}, almost all of which focuses on the three-body interactions relevant to the simplest of all non-Abelian FQH states: the Moore-Read (MR) state\cite{MR}. In practice, the perturbative calculations are computationally involved, giving individual three-body pseudopotentials (PPs) which only lead to a partial understanding of the full interaction. It is useful to have full analytic expressions of the three-body interactions from the LL mixing of realistic experimental systems, from which different physical contributions can be transparently interpreted. For numerical analysis, such expressions also allow us to perform computations without using truncated PPs, as such truncations can lead to approximations with unpredictable errors when combined with finite size effects\cite{simon,morf,haldane,dassarma2,dassarma3,morf2}. 

From experimental perspectives, there still remain many challenges in the study of the Moore-Read states, even though the plateau of the Hall conductivity at filling factor $\mu=5/2$ has been unambiguously observed\cite{english,west}. The main difficulty is because the MR state is quite fragile, with small incompressibility gap requiring very high quality samples\cite{west2,chi1,chi2,morpurgo}; and there exists ambiguity between the Pfaffian (Pf) and anti-Pfaffian (APf) phases in realistic systems, which belongs to two distinct non-Abelian universality classes at the same filling factor\cite{rosenow,fisher}. The APf is the particle-hole conjugate of Pf, with the same energy for any two-body interactions. Three-body interactions from LL mixing break particle-hole symmetry and could be decisive on which phase is favoured. Recent propositions of the particle-hole Pfaffian (PH-Pf) also lead to more possibilities\cite{son,feldman}. Given the sensitivity of the nature of the MR states to various experimental details, additional theoretical tools with better insights can go a long way in tackling the experimental difficulties in realising robust MR states with unambiguous non-Abelian statistics\cite{rosenow,fisher,bonderson}. 

In this paper, we describe a first quantized formalism in deriving full analytic expressions of effective three-body interactions from LL mixing using the Shrieffer-Wolff (SW) transformation\cite{sw}. The formalism can be applied to anisotropic systems\cite{HaldaneGeometry,suyi,yang1,yang2,yang3,zlatko}, or systems where LLs are not equally spaced (no Galilean invariance\cite{shenyu}). The analytical expressions show clear and illuminating contributions from realistic bare interactions, LL spectrum, form factors within LL, form factors from virtual excitations into higher LLs, and various metrics in the system. We study some general features of the three-body interactions. In particular, we show that small breaking of the Galilean invariance can significantly drive the interaction from favouring APf to Pf phase. Given that in realistic materials Galilean invariance is naturally broken with higher orders in band dispersion, this has deep implications on the importance of the detailed band structures for non-Abelian phases at half-filled LLs.

{\it Shrieffer-Wolff Transformation--} As a form of degenerate perturbative calculation, the Shrieffer-Wolff (SW) transformation for quantum many-body systems is a standard and powerful tool for a number of applications\cite{sw}. Starting with the full Hamiltonian of the quantum Hall system as follows:
\begin{eqnarray}
H&=&\sum_ih\left(a_i^\dagger a_i\right)+\int d^2qV_{\vec q}\sum_{i\neq j}e^{iq_ar_i^a}e^{-iq_ar_j^a}\label{part1}\\
&=&H_0+H_1+H_2\label{part2}
\end{eqnarray}
where $H_0=\sum_ih\left(a_i^\dagger a_i\right)$ is the kinetic energy with $a_i,a_i^\dagger$ as ladder operators defining the LLs, and $h\left(x\right)$ can be non-linear for systems without Galilean invariance; $i,j$ are particle indices. In Eq.(\ref{part1}) the interaction part comes form the fundamental two-body Coulomb interaction between electrons. For the special case of two-dimensional systems with zero thickness, $V_{\vec q}=1/q$ with $q=|q|$, the Fourier component of the Coulomb interaction. We focus on the perturbative limit that the energy scale of $H_0$ dominates. The complication comes from the fact that while $H_0$ is diagonal in LLs, it does not commute with the interaction part of $H$: the Coulomb interaction will mix LLs. We thus separate the interaction energy into two parts ($H_1, H_2$) in Eq.(\ref{part2}), with $[H_0,H_1]=0$. Explicitly we have
\begin{eqnarray}\label{h2}
&&H_2=\int d^2qV_{\vec q}e^{-q^2/2}\sum_{i\neq j}f_q\left(R_{ij}\right)\sum'c_{1234}^q\hat V_{ij}^{1234}\\
&&\hat V_{ij}^{1234}=\left(a_i^\dagger\right)^{n_1}\left(a_i\right)^{n_3}\left(a_j^\dagger\right)^{n_2}\left(a_j\right)^{n_4}\\
&&f_q\left(R_{ij}\right)=e^{iq_xR_{ij}^x+iq_yR_{ij}^y}\\
&&c_{1234}^q=\frac{\left(i\textbf{q}\right)^{n_1}\left(-i\textbf{q}\right)^{n_2}\left(i\textbf{q}^*\right)^{n_3}\left(-i\textbf{q}^*\right)^{n_4}}{n_1!n_2!n_3!n_4!}
\end{eqnarray}
where $\textbf{q}=\left(q_x+iq_y\right)/\sqrt 2$, $\sum'$ sums over all terms that does not commute with $H_0$, and $R_{ij}^{x,y}=R_i^{x,y}-R_j^{x,y}$ are guiding center coordinates commuting with $a_i,a_i^\dagger$. $H_1$ is identical to $H_2$ with the only difference that the summation goes over all terms commuting with $H_0$. We would like to construct an effective Hamiltonian $H_{\text{eff}}=e^{\mathcal S}He^{-\mathcal S}$ that is diagonal in LLs, related to the original Hamiltonian via an SW transformation with the antiunitary operator $\mathcal S$. We organise $\mathcal S$ and the transformation as follows:
\begin{eqnarray}
\mathcal S=\sum_{n=1}^\infty S_n\label{ss},\quad H_{\text{eff}}=H+[\mathcal S,H]+\frac{1}{2}[\mathcal S,[\mathcal S,H]]+\cdots\label{expansion}
\end{eqnarray}
with $S_n\sim\left(\Delta\right)^{-n}$, which is the inverse power of the typical energy scale in $H_0$. If LLs are equally spaced, $\Delta^{-1}\sim\kappa=e^2/\left(\hbar\omega_c\epsilon l_B\right)$, the small parameter commonly used in literature\cite{nayak2, rezayi,simon,zaletel,jain,troyer,nayak3,macdonald}. Our goal is to keep $H_{\text{eff}}$ diagonal in LL order by order. To $\mathcal O\left(1\right)$ we have $H_{\text{eff}}=H+[S_1,H_0]+\mathcal O\left(\Delta^{-1}\right)$. We thus need the following condition:
\begin{eqnarray}\label{c1}
H_2+[S_1,H_0]=0
\end{eqnarray}
This can be solved exactly, with the following explicit expression that can be easily checked by plugging into Eq.(\ref{c1}):
\footnotesize
\begin{eqnarray}\label{s1}
&&S_1=\int d^2qV_qe^{-q^2/2}\sum_{i\neq j}f_q\left(R_{ij}\right)\sum'c^q_{1234}\hat V_{ij}^{1234}\hat G_{ij}^{1234}\\
&&\hat G_{ij}^{1234}=\left(h_i^{n_1n_3}+h_j^{n_2n_4}-h_i^{00}-h_j^{00}\right)^{-1}
\end{eqnarray} 
\normalsize
where we have $h_i^{mn}=h\left(a_i^\dagger a_i-n+m\right)$, and $[\hat G_{ij}^{1234},H_0]=0$. So far we obtained $H_{\text{eff}}=H_0+H_1$ in the limit of $\Delta\rightarrow\infty$. To go beyond the lowest order, we note that at $\mathcal O\left(\Delta^{-1}\right)$ we have
\footnotesize
\begin{eqnarray}\label{c2}
&&[S_1,H_1+H_2]+\frac{1}{2}[S_1,[S_1,H_0]]=[S_1,H_1+\frac{1}{2}H_2]
\end{eqnarray}
\normalsize
where we have used Eq.(\ref{c1}), and Eq.(\ref{c2}) again contains a part that commutes with $H_0$. Just like we used $S_1$ to cancel the part that does not commute with $H_0$ at $\mathcal O\left(1\right)$, we now assume that we can use $S_2$ to cancel the part that does not commute with $H_0$ at $\mathcal O\left(\Delta^{-1}\right)$. We thus only need to calculate the LL conserving part of Eq.(\ref{c2}), given the explicit expression in Eq.(\ref{s1}). This part contains corrections to both two-body and three-body interactions from LL mixing at the order of $\Delta^{-1}$.

{\it Effective Three-body Interactions--} To extract from Eq.(\ref{c2}) the part that commutes with $H_0$, one should note that given $[S_1,H_0]\neq 0$ and $[H_1,H_0]=0$, we only need to compute $[S_1,H_2]$. With some straightforward algebra\cite{footnote1}, both the two-body and three-body interactions can be explicitly evaluated. In this work, we only focus on the three-body interactions, which in the LLL is given as follows:
\begin{eqnarray}
&&V^{\left(0\right)}_{\text{3bdy}}\left(\vec q_1,\vec q_2\right)=V_{\vec q_1}V_{\vec q_2}\mathcal J_0\left(\frac{q_1^2}{2},\frac{q_2^2}{2}\right)\mathcal F_0\left(\vec q_1,\vec q_2\right)\label{lll3}\\
&&\mathcal J_\alpha\left(x,y\right)=e^{-x-y}L_\alpha\left(x\right)L_\alpha\left(y\right)\\
&&\mathcal F_0\left(\vec q_1,\vec q_2\right)=-\sum_{n=1}^\infty\frac{\left(-P_{12}\right)^n}{\Delta_0^nn!}\cos\left(n\theta_{12}+Q_{12}\right)\label{form0}
\end{eqnarray}
Here $q_i$ is the magnitude of $\vec q_i$, $\Delta_m^n=h\left(n\right)-h\left(m\right)$ is the energy difference between LLs; $\theta_{12}=\theta_2-\theta_1$ is the angle between $\vec q_1$ and $\vec q_2$; $L_k\left(x\right)$ is the $k^{\text{th}}$ Laguerre polynomial. We also defined $P_{12}=\frac{1}{2}q_1q_2$ and $Q_{12}=\frac{1}{2}|\vec q_1\times\vec q_2|$. The form factor from virtual excitations in Eq.(\ref{form0}) is a well-behaved function that can be very well approximated by keeping a few terms in the summation. It also has an equivalent compact form expressed in terms of incomplete Gamma functions when we have Galilean invariance\cite{footnote1}, though the physics is more transparent with the explicit sum. One can also easily introduce non-trivial metrics into Eq.(\ref{lll3}) for systems with anisotropy, e.g. with phosphorene systems\cite{phosphorene}.

The three-body interaction in the first LL (1LL) can also be readily derived as follows:
\begin{eqnarray}
&&V^{\left(1\right)}_{\text{3bdy}}\left(\vec q_1,\vec q_2\right)=V_{\vec q_1}V_{\vec q_2}\mathcal J_1\left(\frac{q_1^2}{2},\frac{q_2^2}{2}\right)\mathcal F_1\left(\vec q_1,\vec q_2\right)\label{1ll3}\\
&&\mathcal F_1\left(\vec q_1,\vec q_2\right)=-\frac{P_{12}}{\Delta^1_0}\cos\left(\theta_{12}-Q_{12}\right)\nonumber\\
&&-\sum_{n=1}^\infty\frac{\left(-P_{12}\right)^n}{\Delta^{n+1}_1n!}\cos\left(n\theta_{12}+Q_{12}\right)L^{\left(n\right)}_{12}\label{form1}\\
&&L^{\left(n\right)}_{12}=n+\frac{P_{12}^2}{n+1}+\frac{1}{2}\left(L_1\left(q_1^2\right)+L_1\left(q_2^2\right)\right)
\end{eqnarray}
\normalsize
We can readily verify the correctness of Eq.(\ref{lll3}) and Eq.(\ref{1ll3}) by evaluating the three-body PPs via calculating the energy expectations of antisymmetric three-particle wavefunctions in LLL and 1LL\cite{laughlin}. We have thoroughly and rigorously performed the checking both for three-body and two-body interaction corrections from LL mixing at $\mathcal O\left(\Delta^{-1}\right)$; the results agree perfectly with those reported in the literature if we take $h\left(x\right)=\hbar\omega_c \left(x+1/2\right)$, where $\omega_c$ is the cyclotron frequency\cite{macdonald,nayak3,footnote1}. In addition, Eq.(\ref{lll3})$\sim$Eq.(\ref{1ll3}) are valid for any types of bare interactions $V_{\vec q}$, both for isotropic and anisotropic systems. They are also valid for any LL spectrum, which is useful for graphene like systems\cite{footnote2}.

\begin{figure}[htb]
\includegraphics[width=8cm]{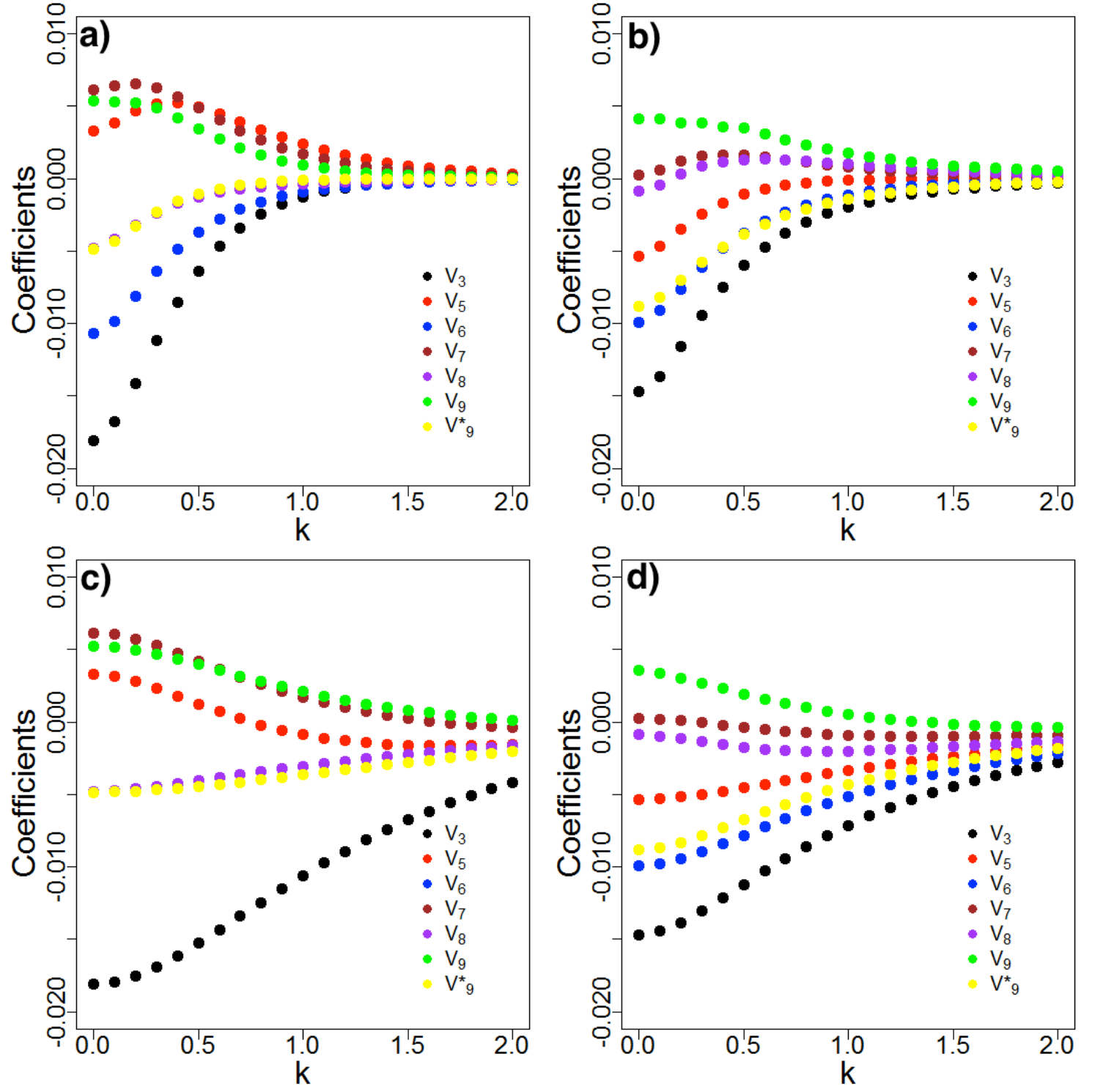}
\caption{Evolution of three-body PPs as a function of the softening parameter $k$. a). $V_{\vec q}=1/\sqrt{q^2+k^2}$ in LLL; b). $V_{\vec q}=1/\sqrt{q^2+k^2}$ in 1LL; c). $V_{\vec q}=e^{-kq}/q$ in LLL; d). $V_{\vec q}=e^{-kq}/q$ in 1LL.}
\label{fig1}
\end{figure}

{\it Realistic Interactions--} We first look at Galilean invariant systems with $h\left(x\right)=\hbar\omega_c \left(x+1/2\right)$ so that LLs are equally spaced. We can now replace $V_{\vec q}$ in Eq.(\ref{lll3}) and Eq.(\ref{1ll3}) with various different experimentally realizable bare interactions. With the explicit three-body interactions, it is very convenient to explore the behaviours of three-body PPs directly. In Fig.(\ref{fig1}) we take $V_{\vec q}=\frac{1}{q}e^{-kq}$ and $V_{\vec q}=\frac{1}{\sqrt{q^2+k^2}}$ respectively, modelling the 2D electron gas in a sample with a finite thickness\cite{dassarma3}, or with screening\cite{ronny}. One can see that in both cases, increasing the sample thickness or screening reduce the effects of LL mixing. It is thus a general trend that LL mixing is suppressed when the Coulomb interaction is softened one way or another.

For Galilean invariant, isotropic systems, we analysed a large class of interactions, and found that as long as the bare interaction $V_{\vec q}$ decays for large $q$ monotonically, $V_3$ is generally the dominant three-body PP. The sign of $V_3$ is of particular importance for the MR state at half-filling. This is because in addition to breaking the PH symmetry, a positive $V_3$ gives the model Hamiltonian for the Pf state (of which it is the exact zero energy state), and a negative $V_3$ plus a specific two-body interaction\cite{phdassarma} is the model Hamiltonian for the APf state. All the interactions we analysed have negative $V_3$, and all PPs become weaker when the bare interaction is softened (e.g. by the finite thickness). We thus expect that in realistic systems, the non-Abelian topological phase is mostly stabilized by the effective two-body interactions. However, small three-body interactions from LL mixing could be crucial in determining if we have Pf,APf or PH-Pf phases, or even a mixture of two phases with domain walls\cite{son,feldman,simon2,defeldman,heiblum,halperin2,wan}. Given that all varieties of bare interactions we have analysed give negative $V_3$, this is strong evidence that in Galilean invariant, isotropic experimental systems, the Pf is \emph{not} the preferred MR state, even with realistic interactions having details not exactly captured in our analysis (e.g. detailed confining potentials in the vertical direction). 

{\it Non-Galilean invariant systems--} Apart from degrees of freedom such as spins, different experimental systems are fully characterized by the bare interaction profile $V_{\vec q}$ (which could be anisotropic relative to the effective mass tensor), and $\Delta_m^n$ intrinsically determined by $h\left(x\right)$, the single particle LL spectrum. In general, Galilean invariance is only an approximation in real materials, when the Fermi level is near the bottom (top) of the conduction (valence) band\cite{footnote1}. Higher order terms in momenta generally exist, however, leading to LL spectrums not equally spaced. 

It is worth noting from Eq.({\ref{lll3}) and Eq.(\ref{1ll3}) that virtual excitations into different LLs are suppressed in two ways: the factorial suppression from the Taylor expansion of the density operators, and the suppression due to the energy gaps between LLs. The latter can in principle be tuned in experiments, so it is instructive to first look at the ``bare contribution" from the virtual excitations to each LL, without the suppression of the LL energy gaps. As one can see from Fig.(\ref{fig2}), even without energy suppressions, bare contributions from higher LLs decreases rapidly with LL index. In the LLL, most of the effects come from virtual excitations into the 1LL, which gives a large negative $V_3$ contribution. As a result, we do not expect non-linearity in $h\left(x\right)$ could overcome this dominant contribution. We thus expect Pf to be always disfavoured in the LLL (if an incompressible phase exists) when other factors are ignored.
\begin{figure}[htb]
\includegraphics[width=8cm]{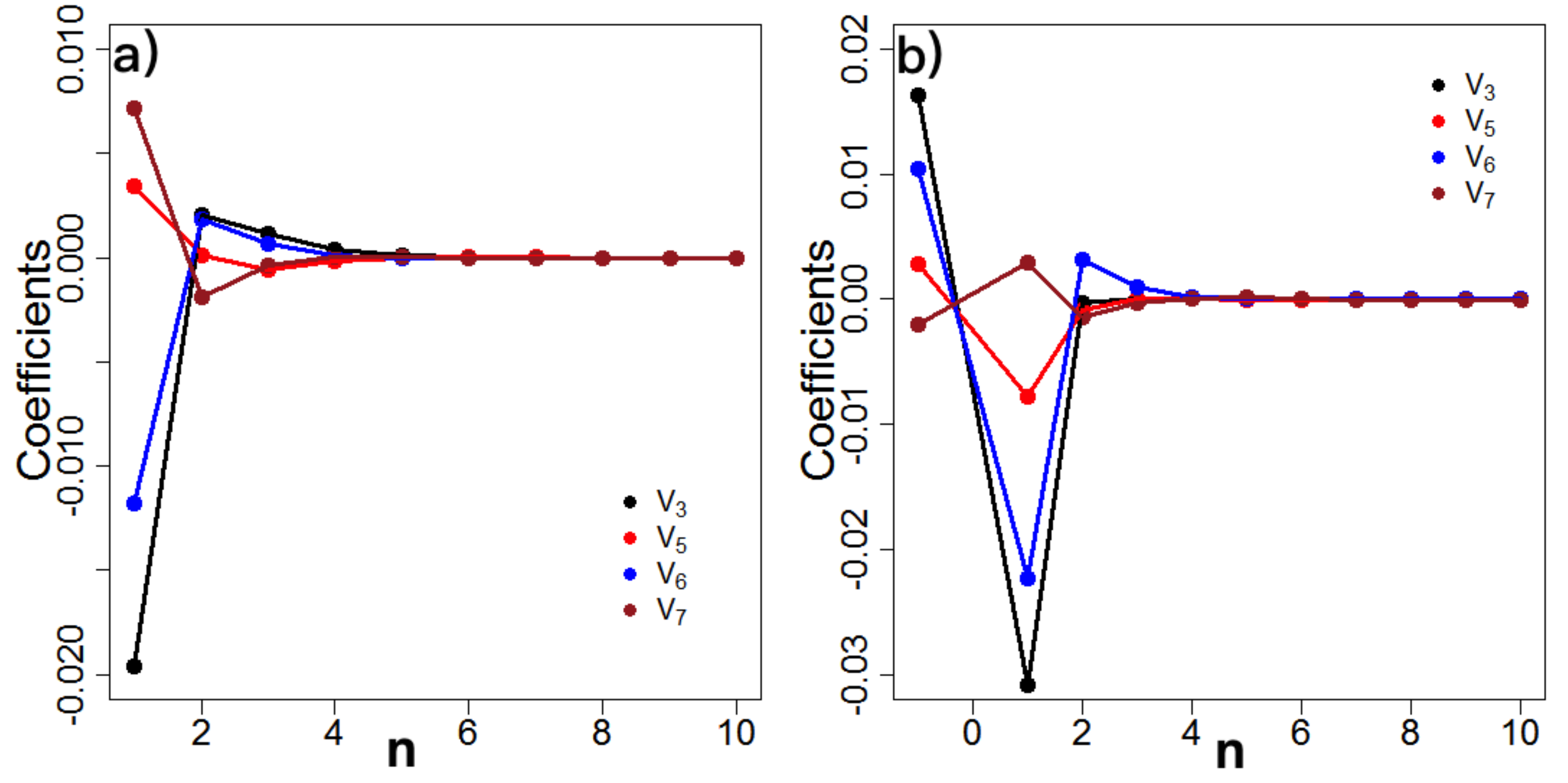}
\caption{Contributions from each LL to different three-body PPs, computed from Eq.(\ref{lll3}) and Eq.(\ref{1ll3}), without the energy gap suppressions in Eq.(\ref{form0}) and Eq.(\ref{form1}). Only one term in the summation in Eq.(\ref{form0}) or Eq.(\ref{form1}) is used, indexed by $n$ which is the x-axis.  a). Contributions to effective interactions in LLL; b). Contributions to effective interactions in 1LL, $n=-1$ is given by virtual excitations to and from LLL.}
\label{fig2}
\end{figure} 

It is interesting to see that in the 1LL, the virtual excitation to and from the LLL gives a positive contribution to $V_3$, while those to all higher LLs give negative contributions. We can thus in principle tune the LL spacing to enhance the suppression into higher LLs, to bring the overall $V_3$ from negative to positive. In Fig.(\ref{fig3}a) we plot various three-body PPs with the simple model of $h\left(x\right)=x+1/2+k\left(x+1/2\right)^4$, with Coulomb interaction $V_{\vec q}=1/q$. While clearly we have a dominant negative $V_3$ at $k=0$, for very small non-linearity there is a transition to a dominant positive $V_3$. If the two-body interactions support a sizeable incompressibility gap at half filling, we take this as strong evidence of a possible APf to Pf transition, driven by the breaking of Galilean invariance, or the higher orders of the band structure. In Fig.(\ref{fig3}b) we look at $h\left(x\right)=x+1/2+k_1\left(x+1/2\right)^2+k_2\left(x+1/2\right)^3$, and plot the relative strength of $V_3$ as a function of $\Delta_0^1,\Delta_1^2$ (uniquely determined by $k_1,k_2$). Contributions from higher LLs can be safely ignored based on Fig.(\ref{fig2}), and one can see possible regions where Pf or APf phases can be stabilized by LL mixing. More extensive numerical analysis are needed to further investigate such phase transitions, especially for systems with disorder, particularly with the recent proposal of particle-hole Pfaffian (PH-Pf)\cite{son,feldman}. 
 \begin{figure}[htb]
\includegraphics[width=9cm]{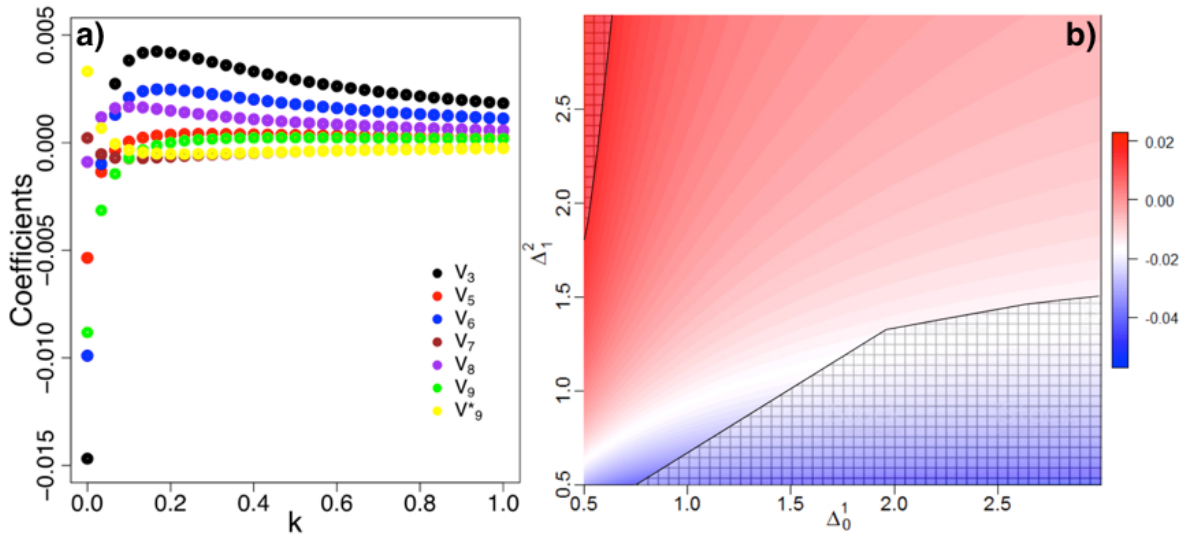}
\caption{In 1LL with $V_{\vec{q}}=1/q$, a). Three-body PPs with $h\left(x\right)=x+1/2+k\left(x+1/2\right)^4$; b). $V_3$ with $\text{1LL}\rightarrow\text{LLL}$ and $\text{1LL}\rightarrow\text{2LL}$ contributions, with different energy gaps. All other LL contributions are ignored because they are small. The two shaded regions are $V_3>0.016$ and $|V_3/V_6|>1.3$, where $V_6$ is the next largest PP. The numbers are chosen arbitrarily to tentatively indicate the possibility of stabilizing Pf and APf phases.}
\label{fig3}
\end{figure}

{\it Summary and Discussions--} In this work, we first show that effective three-body interactions from LL mixing have illuminating analytic expressions that can be readily derived from Shrieffer-Wolff transformation in the first quantized form. This allows us to get around sophisticated perturbative Feynman diagram computations that are more suitable for computing individual PPs. Moreover, the analytic expressions can be used for a wide variety of physical systems, both with or without rotational or Galilean invariance. Different theoretical models or experimental systems are completely characterised by the single particle LL spectrum and the bare interactions between electrons, which can be easily plugged into Eq.(\ref{lll3}) or Eq.(\ref{1ll3}). While we only focus on three-body interactions in the lowest two LLs, the formalism works for all LLs and for effective two-body interaction corrections to the leading order of $\Delta^{-1}$ as well.

Both Eq.(\ref{lll3}) and Eq.(\ref{1ll3}) could be useful for numerical computations especially on torus geometry. We also found in general three-body interactions from LL mixing are strongest for infinitesimally thin, un-screened Coulomb interaction. Softening the Coulomb interaction tends to significantly suppress three-body interaction. Thus experimentally for MR states at $\nu=1/2$ or $5/2$, most probably the incompressibility gaps of Pf or APf are dominated by two-body interactions, at least for cases where spin or valley degrees of freedom are unimportant. Nevertheless, for experimental realizations of robust MR states, the breaking of particle-hole symmetry by LL mixing plays an important role. If the three-body interaction only weakly prefers one phase (Pf or APf) over the other, the experimental systems may have complicated arrangements of Pf and APf domains, or PH-Pf phases due to various factors such as disorder\cite{son,feldman,simon2,defeldman,heiblum,halperin2,wan}. We find that for clean, Galilean invariant systems, physical interactions generally favours the APf phase, though the suppression of the three-body interactions by softening of the Coulomb interaction (very common in experimental systems) can easily make the symmetry breaking ambiguous, which may explain some of the experimental difficulties in realising very robust APf or Pf phases\cite{stern}.

We also particularly explored systems without Galilean invariance, which is a common feature in most materials, though not very well investigated before. When the Fermi level moves away from the bottom (top) of the conduction (valence) band, higher orders terms (beyond the quadratic term) of the band dispersion are generally present, leading to LLs not equally spaced. Such physics can be explicitly captured in the analytical expressions of three-body interactions we derived. Most interestingly, in 1LL (and higher LLs in general, but not in LLL), we found that higher order terms in the band dispersion can drive the three-body interactions from being dominated by negative $V_3$ to being dominated by positive $V_3$. As long as the Coulomb interaction is not overly softened, this will very possibly lead to a phase transition from APf to Pf via band engineering. We thus propose that detailed band dispersions in the experimental systems deserve more attentions(even for cases where quadratic dispersion dominates) for the MR state, because they could be essential ingredients in determining the nature of the topological phases found in experiments at $\nu=5/2$.

The formalism we have derived is in the limit of zero disorder (clean, translationally invariant systems), while it is recently pointed out\cite{feldman,heiblum,halperin2} that disorder could play an important role for stabilising the PH-Pf state. Weak disorder (as compared to electron-electron interaction) that does not mix LL can be added in a straightforward way as one-body orbital potentials to the effective Hamiltonian within a single LL. For relatively strong disorder such that the induced LL mixing becomes important, more perturbative computations are needed. The study of interplay between LL mixing and correlated disorders in realistic samples could be crucial for the complete understanding of the topological phases at $\mu=5/2$.

{\sl Acknowledgements.} I thank Wen Wei Ho from Harvard University for directing my attention to SW transformation, as well as useful discussions with Ching Hua Lee from Institute of High Performance Computing of Singapore, and Zi-Xiang Hu from Chongqing University. This work is supported by the NTU grant for Nanyang Assistant Professorship.

\begin{center}
\textbf{\large Supplemental Online Material for ``Aspects of Three-body Interactions in Generic Fractional Quantum Hall Systems and Impact of Galilean Invariance Breaking" }\\[5pt]
\begin{quote}
{\small In this supplementary material we include analytical details on the derivation of the full analytic expressions for the effective three-body interactions in the lowest Landau level (LLL) and the first Landau level (1LL), the verifications of these expressions by computing the three-body pseudopotentials (PPs) and comparing with known values in the literature, as well as explanations of Galilean invariance breaking in real materials.}\\[20pt]
\end{quote}
\end{center}
\setcounter{equation}{0}
\setcounter{figure}{0}
\setcounter{table}{0}
\setcounter{page}{1}
\setcounter{section}{0}
\makeatletter
\renewcommand{\theequation}{S\arabic{equation}}
\renewcommand{\thefigure}{S\arabic{figure}}
\renewcommand{\thesection}{S\Roman{section}}
\renewcommand{\thepage}{S\arabic{page}}
\vspace{1cm}
\twocolumngrid

\section{S1. Detailed derivation of effective three-body interactions}\label{derivation}
For a two-dimensional system with a strong perpendicular magnetic field, the canonical momentum is given by $\pi_a=P_a-eA_a$, where $P_a=-i\partial_a$ (we set $\hbar=1$) is the momentum operator, $A_a$ is the vector potential with $\epsilon^{ab}\partial_aA_b=B$, the magnetic field. Einstein's summation rule is implied and $a,b=x,y$ are the spatial directions, $e$ is the particle charge. The magnetic field defines the magnetic length $l_B=1/\sqrt{eB}$, thus the spatial coordinates $r^a$ can now be separated into the cyclotron coordinates $\tilde R^a$ and the guiding center coordinates $R^a$ as follows:
\begin{eqnarray}\label{coordinates}
&&r^a=\tilde R^a+R^a, \qquad\tilde R^a=l_B^2\epsilon^{ab}\pi_b\label{r1}\\
&&[\tilde R^a,\tilde R^b]=il_B^2\epsilon^{ab},\quad[R^a,R^b]=-il_B^2\epsilon^{ab}\label{r2}\\
&&[\tilde R^a,R^b]=0
\end{eqnarray}
We can thus define two sets of decoupled ladder operator operators $a=\left(\tilde R^x+i\tilde R^y\right)/\left(l_B\sqrt 2\right), b=\left(R^x-iR^y\right)/\left(l_B\sqrt 2\right)$ with $[a,a^\dagger]=[b,b^\dagger]=1,[a,b]=[a,b^\dagger]=0$. In particular, $a,a^\dagger$ move particles in between different Landau Levels (LL), while $b,b^\dagger$ move particles within the same LL.

The most generic Hamiltonian for the quantum Hall system is thus given by
\begin{eqnarray}\label{h}
H&=&\sum_ih\left(a_i^\dagger a_i\right)+\int d^2qV_{\vec q}\sum_{i\neq j}e^{iq_ar_i^a}e^{-iq_ar_j^a}
\end{eqnarray}
where $i$ is the particle index, $h$ is the single particle kinetic energy defining the LLs, and $V_{\vec q}$ is the bare interaction between particles. For the special case of two-dimensional systems with zero thickness, $V_{\vec q}=1/q$ with $q=|q|$, the Fourier component of the Coulomb interaction. The second term in Eq.(\ref{h}) is the usual two-body density-density interaction.

We now divide the full Hamiltonian into three parts as follows:
\footnotesize
\begin{eqnarray}
&&H=H_0+H_1+H_2\label{hh}\\
&&H_0=\sum_ih\left(a_i^\dagger a_i\right)\\
&&H_1=\int d^2qV_{\vec q}e^{-q^2/2}\sum_{i\neq j}f_q\left(R_{ij}\right)\sum''c_{1234}^q\hat V_{ij}^{1234}\label{h1}\\
&&H_2=\int d^2qV_{\vec q}e^{-q^2/2}\sum_{i\neq j}f_q\left(R_{ij}\right)\sum'c_{1234}^q\hat V_{ij}^{1234}\label{h2}\\
&&\hat V_{ij}^{1234}=\left(a_i^\dagger\right)^{n_1}\left(a_i\right)^{n_3}\left(a_j^\dagger\right)^{n_2}\left(a_j\right)^{n_4}\\
&&f_q\left(R_{ij}\right)=e^{iq_xR_{ij}^x+iq_yR_{ij}^y}\\
&&c_{1234}^q=\frac{\left(i\textbf{q}\right)^{n_1}\left(-i\textbf{q}\right)^{n_2}\left(i\textbf{q}^*\right)^{n_3}\left(-i\textbf{q}^*\right)^{n_4}}{n_1!n_2!n_3!n_4!}\label{cq}
\end{eqnarray}
\normalsize
where $H_0$ is the kinetic energy, the interaction is separated into two parts $H_1, H_2$, in which we used Eq.(\ref{r1}) and expanded the exponentials containing $a_i,a_i^\dagger$. We also define $\textbf{q}=\left(q_x+iq_y\right)/\sqrt 2$ and $R_{ij}^{x,y}=R_i^{x,y}-R_j^{x,y}$ (note they commute with $a_i,a_i^\dagger$).

In Eq.(\ref{h1}) the summation $\sum''$ goes over all terms such that $[H_0,\hat V_{ij}^{1234}]=0$; in contrast in Eq.(\ref{h2}) the summation $\sum'$ goes over all terms such that $[H_0,\hat V_{ij}^{1234}]\neq 0]$. The separation of the interaction into two parts thus gives $[H_0,H_1]=0,[H_0,H_2]\neq 0$. One should also note the following identity:
\begin{eqnarray}\label{e1}
&&[H_0,\hat V_{ij}^{1234}]=\hat V_{ij}^{1234}\hat F_{ij}^{1234}\label{c1}\\
&&\hat F_{ij}^{1234}=h\left(a_i^\dagger a_i-n_3+n_1\right)+h\left(a_j^\dagger a_j-n_4+n_2\right)\nonumber\\
&&\qquad\qquad\qquad-h\left(a_i^\dagger a_i\right)-h\left(a_j^\dagger a_j\right)
\end{eqnarray}
In cases that $[H_0,\hat V_{ij}^{1234}]\neq 0$, $\hat F_{ij}^{1234}$ is invertible, and we can define 
\begin{eqnarray}\label{g}
\hat G_{ij}^{1234}=\left(\hat F_{ij}^{1234}\right)^{-1}.
\end{eqnarray}
and we are ready to implement the Schrieffer-Wolff transformation.

\subsection{Schrieffer-Wolff Transformation}

In Eq.(\ref{hh}), the only term that mixes LLs is $H_2$, and our goal is to find an antiunitary operator $\mathcal S$ such that 
\begin{eqnarray}\label{transform}
H_{\text{eff}}&=&e^{\mathcal S}He^{-\mathcal S}\\
&=&H+[\mathcal S,H]+\frac{1}{2}[\mathcal S,[\mathcal S,H]]+\cdots
\end{eqnarray}
resulting in $H_{\text{eff}}, H$ having identical energy spectrum, but $H_{\text{eff}}$ does not mix LLs. As a degenerate perturbative scheme, Schrieffer-Wolff (SW) transformation is only valid in the limit that the energy scale of $H_0$ dominates, which we assume throughout this work. We thus organize the antiunitary operator as follows:
\begin{eqnarray}\label{s}
\mathcal S=\sum_nS_n,\qquad S_n\sim\mathcal O\left(\Delta^{-n}\right)
\end{eqnarray}
where $\Delta$ is the energy scale of $H_0$ (we take the energy scales of $H_1,H_2$ as $\mathcal O\left(1\right)$). The strategy is to compute $H_{\text{eff}}$ in Eq.(\ref{transform}) order by order. The leading order is obviously $H_0$ at $\mathcal O\left(\Delta\right)$, which does not mix LLs. At $\mathcal O\left(1\right)$ we have
\begin{eqnarray}\label{o1}
H_{\text{eff}}=H_0+H_1+H_2+[S_1,H_0]+\mathcal O\left(\Delta^{-1}\right)
\end{eqnarray}
Clearly the necessary and sufficient condition here is to have
\begin{eqnarray}\label{s1h0}
[S_1,H_0]=-H_2.
\end{eqnarray}
Using Eq.(\ref{h2}) and Eq.(\ref{c1}), Eq.(\ref{s1h0}) can be explicitly solved with the following expression:
\footnotesize
\begin{eqnarray}\label{ss1}
S_1=\int d^2qV_qe^{-q^2/2}\sum_{i\neq j}f_q\left(R_{ij}\right)\sum'c^q_{1234}\hat V_{ij}^{1234}\hat G_{ij}^{1234}
\end{eqnarray}
\normalsize
Note that $S_1$ only contains terms that mix LLs, and $[\hat G_{ij}^{1234},H_0]=0$. One can obtain other solutions to Eq.(\ref{s1h0}) by adding terms to $S_1$ that commutes with $H_0$, this will not affect the results derived in this work. The resulting effective Hamiltonian $H_{\text{eff}}=H_0+H_1+\mathcal O\left(\Delta^{-1}\right)$ is the usual effective two-body interaction with no LL mixing, where $H_0$ can be treated as a constant if all particles are within a single LL.

We now move onto $\mathcal O\left(\Delta^{-1}\right)$, which is the order of interest in this work. Writing $H_{\text{eff}}^{\left(0\right)}=H_0,H_{\text{eff}}^{\left(1\right)}=H_1$, all relevant terms at $\mathcal O\left(\Delta^{-1}\right)$ are given by
\begin{eqnarray}\label{o2}
H_{\text{eff}}^{\left(2\right)}=[S_2,H_0]+[S_1,H_1+H_2]+\frac{1}{2}[S_1,[S_1,H_0]]
\end{eqnarray}
Here, we assume all terms in Eq.(\ref{o2}) that will mix LL will be cancelled out by a judicious choice of $S_2$ that does not commute with $H_0$; the first term thus only mixes LL. We do not need to worry about $S_2$ at this order, and only need to derive all terms in Eq.(\ref{o2}) that does not mix LL. Given Eq.(\ref{s1h0}) and that $[S_1,H_1]$ mixes LL (since $S_1$ mixes LL but $H_1$ does not mix LL), the only contributions to $H_{\text{eff}}$ at $\mathcal O\left(\Delta^{-1}\right)$ are as follows:
\begin{eqnarray}
H_{\text{eff}}^{\left(2\right)}=\frac{1}{2}[S_1,H_2]^*
\end{eqnarray}
which we can explicitly evaluate with Eq.(\ref{h2}) and Eq.(\ref{ss1}). The $*$ superscript indicates that only terms that do not mix LLs are included. Explicitly we have
\footnotesize
\begin{eqnarray}\label{s1h1}
[S_1,H_2]=&&\int d^2q_1d^2q_2V_{q_1}V_{q_2}e^{-q_1^2/2-q_2^2/2}\sum_{i\neq j,k\neq l}\sum'c^{q_1}_{1234}c^{q_2}_{5678}\nonumber\\
&&[f_{q_1}\left(R_{ij}\right)\hat V_{ij}^{1234}\hat G_{ij}^{1234},f_{q_2}\left(R_{kl}\right)\hat V_{kl}^{5678}]
\end{eqnarray}
\normalsize
A useful identity for the evaluation we will use is that given $[\hat A_i,\hat B_j]=0$, we have
\begin{eqnarray}\label{e2}
[\hat A_1\hat B_1,\hat A_2\hat B_2]=\hat A_1\hat A_2\{\hat B_1,\hat B_2\}-\{\hat A_1,\hat A_2\}\hat B_2\hat B_1
\end{eqnarray}
A rather trivial reorganization of the commutator in Eq.(\ref{s1h1}) gives the following result:
\begin{widetext}
\begin{eqnarray}\label{s1h1a}
[S_1,H_2]&=&\int d^2q_1d^2q_2V_{q_1}V_{q_2}e^{-q_1^2/2-q_2^2/2}\sum_{i\neq j,k\neq l}\sum'c^{q_1}_{1234}c^{q_2}_{5678}f_{q_1}\left(R_{ij}\right)f_{q_2}\left(R_{kl}\right)\nonumber\\
&&\hat V_{ij}^{1234}\left(\hat G_{ij}^{1234}\hat V_{kl}^{5678}-\hat V_{kl}^{5678}\hat G_{kl}^{5678}\right)
\end{eqnarray}
\end{widetext}
In Eq.(\ref{s1h1a}), there are both terms that connect particles in different LLs, and terms that only connect particles in the same LLs (after virtual excitations into other LLs). We are only interested in the latter, as the former will be cancelled by $[S_2,H_0]$ in Eq.(\ref{o2}). Thus the only possibilities are (modulus permutation of particle indices): a). $i=k,j=l$, which leads to effective two-body interactions at $\mathcal O\left(\Delta^{-1}\right)$; b). $i=k,j\neq l$, which leads to effective three-body interactions at $\mathcal O\left(\Delta^{-1}\right)$. We will now evaluate the effective three-body interactions in LLL and 1LL respectively.

\subsection{Three-body interactions in LLL}

We now look at the case of $i=k, j\neq l$ for the derivation of the effective three-body interaction (indices permutations give identical results). In LLL, the only non-zero contributions in Eq.(\ref{s1h1a}) come from $n_1=n_2=n_4=n_6=n_7=n_8=0$, and $n_3=n_5>0$. In particular, let $|0\rangle$ be the state where all particles are in LLL, we have the following:
\begin{eqnarray}\label{lll}
&&\langle 0|\hat V_{ij}^{1234}\hat G_{ij}^{1234}\hat V_{il}^{5678}-\hat V_{ij}^{1234}V_{il}^{5678}\hat G_{il}^{5678}|0\rangle\nonumber\\
&&=n_3!\left(\bar G^{00n_30}_{n_30}-\bar G_{00}^{n_3000}\right)\\
&&\left(\bar G_{MN}^{n_1n_2n_3n_4}\right)^{-1}=\Delta_M^{M-n_3+n_1}+\Delta_N^{N-n_4+n_2}
\end{eqnarray}
where $\Delta_m^n$ is the kinetic energy gap between the $m^{\text{th}}$ and $n^{\text{th}}$ LL as defined in the main text. If LLs are equally spaced with gaps between adjacent LLs given by $\hbar\omega_c$, where $\omega_c$ is the cyclotron energy, then $\Delta_m^n=\hbar\omega_c\left(m-n\right)$. We can thus rewrite Eq.(\ref{s1h1a}) as follows:
\begin{eqnarray}
&&[S_1,H_2]^*=\int d^2q_1d^2q_2V_{q_1}V_{q_2}e^{-q_1^2/2-q_2^2/2}\nonumber\\
&&\qquad\qquad\quad\sum_{i\neq j\neq k}e^{iq_{1a}R_i^a}e^{iq_{2a}R_j^a}e^{-i\left(q_{1a}+q_{2a}\right)R_k^a}\\
&&=\int d^2q_1d^2q_2V_{\text{3bdy}}^{\left(0\right)}\sum_{i\neq j\neq k}e^{iq_{1a}R_i^a}e^{iq_{2a}R_j^a}e^{-i\left(q_{1a}+q_{2a}\right)R_k^a}\nonumber\\
\end{eqnarray}
Using Eq.(\ref{cq}) and the corresponding values of $n_1\sim n_8$, we arrive at the results presented in the main text, as we quote here:
\footnotesize
\begin{eqnarray}
&&V^{\left(0\right)}_{\text{3bdy}}\left(\vec q_1,\vec q_2\right)=V_{\vec q_1}V_{\vec q_2}\mathcal J_0\left(\frac{q_1^2}{2},\frac{q_2^2}{2}\right)\mathcal F_0\left(\vec q_1,\vec q_2\right)\label{lll3}\\
&&\mathcal J_\alpha\left(x,y\right)=e^{-x-y}L_\alpha\left(x\right)L_\alpha\left(y\right)\\
&&\mathcal F_0\left(\vec q_1,\vec q_2\right)=-\sum_{n=1}^\infty\frac{\left(-P_{12}\right)^n}{\Delta_0^nn!}\cos\left(n\theta_{12}+Q_{12}\right)\label{form0}
\end{eqnarray}
\normalsize
and we have $\theta_{12}=\theta_2-\theta_1$ as the angle between $\vec q_1$ and $\vec q_2$; $L_k\left(x\right)$ is the $k^{\text{th}}$ Laguerre polynomial. We also defined $P_{12}=\frac{1}{2}q_1q_2$ and $Q_{12}=\frac{1}{2}|\vec q_1\times\vec q_2|=\frac{1}{2}q_1q_2\sin\theta_{12}$. For the special case of Galilean invariance so that LLs are equally spaced, we have $\Delta_0^n=n\hbar\omega_c$, and Eq.(\ref{form0}) is also equivalent to the following expression:
\begin{widetext}
\begin{eqnarray}
\mathcal F_0\left(\vec q_1,\vec q_2\right)=\text{Re}[\left(\gamma+\Gamma\left(0,P_{12}e^{i\theta_{12}}\right)+\log\left(P_{12}e^{i\theta_{12}}\right)\right)e^{iQ_{12}}]
\end{eqnarray}
\end{widetext}
where $\lambda\simeq 0.577216$ is the Euler's constant, and $\Gamma\left(n,x\right)$ is the incomplete Gamma function.

\subsection{Three-body interactions in 1LL}
Similarly starting from Eq.(\ref{s1h1a}), in 1LL the non-zero contributions come from $n_2=n_4=0,1$, $n_6=n_8=0,1$, $n_1,n_7=0,1$. We also have the constraints that $n_1+n_5=n_3+n_7$ and $n_1\neq n_3, n_5\neq n_7$. Let $|1\rangle$ be the state where all particles are in 1LL, we have the following:
\begin{widetext}
\begin{eqnarray}\label{1ll}
&&\langle 1|\hat V_{ij}^{1234}\hat G_{ij}^{1234}\hat V_{il}^{5678}-\hat V_{ij}^{1234}V_{il}^{5678}\hat G_{il}^{5678}|1\rangle\nonumber\\
&=&\sum_{N>0}\left(N+1\right)!\left(\delta_{n_1,0}\delta_{n_3,N}\delta_{n_5,N}\delta_{n_7,0}+\delta_{n_1,1}\delta_{n_3,N+1}\delta_{n_5,N}\delta_{n_7,0}+\delta_{n_1,0}\delta_{n_3,N}\delta_{n_5,N+1}\delta_{n_7,1}+\delta_{n_1,1}\delta_{n_3,N+1}\delta_{n_5,N+1}\delta_{n_7,1}\right)\nonumber\\
&&\left(\delta_{n_2,0}\delta_{n_4,0}+\delta_{n_2,1}\delta_{n_4,1}\right)\left(\delta_{n_6,0}\delta_{n_8,0}+\delta_{n_6,1}\delta_{n_8,1}\right)\left(\bar G_{N+1,1}^{n_1n_2n_3n_4}-\bar G_{11}^{n_5n_6n_7n_8}\right)\nonumber\\
&&+\delta_{n_1,1}\delta_{n_3,0}\delta_{n_5,0}\delta_{n_7,1}\left(\delta_{n_2,0}\delta_{n_4,0}+\delta_{n_2,1}\delta_{n_4,1}\right)\left(\delta_{n_6,0}\delta_{n_8,0}+\delta_{n_6,1}\delta_{n_8,1}\right)\left(\bar G_{0,1}^{n_1n_2n_3n_4}-\bar G_{11}^{n_5n_6n_7n_8}\right)
\end{eqnarray}
\end{widetext}
The first term in Eq.(\ref{1ll}) involves virtual excitations into higher LLs, while the second term only involves virtual excitations to and from the 1LL and the LLL. Plugging it back into Eq.(\ref{s1h1a}), and with some book-keeping, we obtain the results from the main text which we quote here:
\footnotesize
\begin{eqnarray}
&&V^{\left(1\right)}_{\text{3bdy}}\left(\vec q_1,\vec q_2\right)=V_{\vec q_1}V_{\vec q_2}\mathcal J_1\left(\frac{q_1^2}{2},\frac{q_2^2}{2}\right)\mathcal F_1\left(\vec q_1,\vec q_2\right)\label{1ll3}\\
&&\mathcal F_1\left(\vec q_1,\vec q_2\right)=-\frac{P_{12}}{\Delta^1_0}\cos\left(\theta_{12}-Q_{12}\right)\nonumber\\
&&-\sum_{n=1}^\infty\frac{\left(-P_{12}\right)^n}{\Delta^{n+1}_1n!}\cos\left(n\theta_{12}+Q_{12}\right)L^{\left(n\right)}_{12}\label{form1}\\
&&L^{\left(n\right)}_{12}=n+\frac{P_{12}^2}{n+1}+\frac{1}{2}\left(L_1\left(q_1^2\right)+L_1\left(q_2^2\right)\right)
\end{eqnarray}
\normalsize
Again for the special case of Galilean invariance so that LLs are equally spaced, we have $\Delta_1^{n+1}=n\hbar\omega_c, \Delta_0^1=\hbar\omega_c$, and Eq.(\ref{form1}) is also equivalent to the following expression:
\begin{widetext}
\begin{eqnarray}
\mathcal F_1\left(\vec q_1,\vec q_2\right)=\text{Re}[\left(1-e^{-P_{12}e^{i\theta_{12}}}\left(1+P_{12}e^{-i\theta_{12}}\right)-\left(P_{12}\right)^2+L_1\left(\frac{1}{2}q_1^2\right)L_1\left(\frac{1}{2}q_2^2\right)\mathcal F_0\left(\vec q_1,\vec q_2\right)\right)e^{iQ_{12}}]
\end{eqnarray}
\end{widetext}

\section{S2. Computation of three-body Pseudopotentials from Analytic expressions}\label{computation}
For effective three-body interaction Hamiltonians in a single LL, the most general form is given as follows:
\footnotesize
\begin{eqnarray}\label{3bdy}
\mathcal H=\int d^2q_1d^2q_2V_{\vec q_1\vec q_2}\sum_{i\neq j\neq k}e^{iq_{1a}R_i^a}e^{iq_{2a}R_j^a}e^{-i\left(q_{1a}+q_{2a}\right)R_k^a}
\end{eqnarray}
\normalsize
We use the following preferred coordinates for the Hilbert space of three particles:
\begin{eqnarray}\label{coordinates}
R^a_{ij}&=&\frac{1}{\sqrt{2}}\left(R^a_i-R^a_j\right)\\
R^a_{ij,k}&=&\frac{1}{\sqrt{6}}\left(R^a_i+R^a_j-2R^a_k\right)\\
R^a_{ijk}&=&\frac{1}{\sqrt{3}}\left(R^a_i+R^a_j+R^a_k\right)\label{cm}
\end{eqnarray}
The three sets of coordinates commute with each other. Since we are looking at the case where Eq.(\ref{3bdy}) is translationally invariant, the center of mass coordinates in Eq.(\ref{cm}) do not appear in the Hamiltonian. It is sufficient to look at the case with only three particles, so we set $i=1,j=2,k=3$. We can thus define two sets of ladder operators as follows:
\begin{eqnarray}
&&b_1=\frac{1}{\sqrt 2l_B}\left(R^x_{12}-iR^y_{12}\right)\\
&&b_2=\frac{1}{\sqrt 2l_B}\left(R^x_{12,3}-iR^y_{12,3}\right)\\
&&[b_1,b_1^\dagger]=[b_2,b_2^\dagger]=1,[b_1,b_2]=[b_1,b_2^\dagger]=0
\end{eqnarray}
By defining $\vec q=\left(\vec q_1-\vec q_2\right)/\sqrt 2, \vec {q'}=\sqrt 3\left(\vec q_1+\vec q_2\right)/\sqrt 2$, the Hamiltonian can be rewritten as follows:
\begin{eqnarray}
\mathcal H=\int d^2q_1d^2q_2V_{\vec q_1\vec q_2}e^{i q_a R_{12}^a}e^{i q_a' R_{12,3}^a}
\end{eqnarray}
The Hilbert space of three particles (for now treating them as distinct particles) is indexed by two non-negative integers $|m,n\rangle=\left(b_1^\dagger\right)^m\left(b_2^\dagger\right)^n|0\rangle,b_1|0\rangle=b_2|0\rangle=0$. Similar to the two-body cases, we have the following identity for the matrix elements:
\begin{eqnarray}\label{melement}
&&\langle m,n|e^{i q_a R_{12}^a}e^{i q_a' R_{12,3}^a}|m',n'\rangle=\sqrt{\frac{m!n!}{m'!n'!}}e^{-\frac{1}{2}\left(|q|^2+|q'|^2\right)}\nonumber\\
&&\quad\left(i\sqrt 2\textbf{q}\right)^{\Delta m}\left(i\sqrt 2\textbf{q}'\right)^{\Delta n}L_m^{\Delta m}\left(|q|^2\right)L_n^{\Delta n}\left(|q'|^2\right)
\end{eqnarray}
with $L_m^n\left(x\right)$ as generalized Laguerre polynomial, and $\textbf{q}=\left(q_x+iq_y\right)/\sqrt 2,\textbf{q}'=\left(q'_x+iq'_y\right)/\sqrt 2$.

One should note, however, that we are dealing with identical particles that are either Fermions or Bosons. In this work we focus on Fermions, so that the three particle states have to be fully antisymmetric, and its Hilbert space is spanned by states constructed by Laughlin\cite{laughlin}. Here, it is useful to express these states in the basis of $|m,n\rangle$, and we illustrate this with spin polarized states, using a set of integers $\left(k,l\right)$ to label these states. The relative orbital angular momentum of these states is given by $\alpha=\left(2k+3l\right)$. The first few states are explicitly presented in Table (\ref{t1}), one can easily check that each of them is fully anti-symmetric.
\begin{table}[H]
\centering
\begin{tabular}{| c | c | c |}
\Xhline{3\arrayrulewidth}
&&\\
$\left(k,l\right)$ &  $\alpha$ & $|\Psi_{kl}\rangle$\\
&&\\
\hline
&&\\
$\left(0,1\right)$ & $3$ & $\frac{1}{2}|3,0\rangle-\frac{\sqrt 3}{2}|1,2\rangle$\\
&&\\
\hline
&&\\
$\left(1,1\right)$& $5$ & $-\frac{\sqrt 5}{4}|5,0\rangle+\frac{1}{2\sqrt 2}|3,2\rangle+\frac{3}{4}|1,4\rangle$\\
&&\\
\hline
&&\\
$\left(0,2\right)$ &$6$ &$\frac{\sqrt 3}{4}|5,1\rangle-\frac{1}{2}\sqrt{\frac{5}{2}}|3,3\rangle+\frac{\sqrt 3}{4}|1,5\rangle$\\
&&\\
\hline
&&\\
$\left(2,1\right)$& $7$ &$-\frac{\sqrt {21}}{8}|7,0\rangle+\frac{1}{8}|5,2\rangle+\frac{\sqrt{15}}{8}|3,4\rangle+\frac{3\sqrt 3}{8}|1,6\rangle$\\
&&\\
\hline
&&\\
$\left(1,2\right)$& $8$ &$\frac{3}{4\sqrt 2}|7,1\rangle-\frac{\sqrt 7}{4\sqrt 2}|5,3\rangle-\frac{\sqrt 7}{4\sqrt 2}|3,5\rangle+\frac{3}{4\sqrt 2}|1,7\rangle$\\
&&\\
\hline
&&\\
$\left(0,3\right)$& $9$ &$\frac{1}{16}|9,0\rangle-\frac{3}{8}|7,2\rangle+\frac{3\sqrt 7}{8\sqrt 2}|5,4\rangle-\frac{\sqrt {21}}{8}|3,6\rangle+\frac{3}{16}|1,8\rangle$\\
&&\\
\hline
&&\\
$\left(3,1\right)$& $9$ &$-\frac{\sqrt{21}}{8}|9,0\rangle+\frac{\sqrt 3}{4\sqrt 2}|5,4\rangle+\frac{1}{2}|3,6\rangle+\frac{\sqrt{21}}{8}|1,8\rangle$\\
&&\\
\hline
\end{tabular}
\caption{Expressions of fully anti-symmetric three-body wavefunctions in the basis of $|m,n\rangle$.}
\label{t1}
\end{table}

It is straightforward to calculate the three-body PPs by evaluating the energy expectation of $|\Psi_{kl}\rangle$, simple implementations of numerical integration in Mathematica with ``MultiDimensionalRule" give sufficiently high accuracy. Using $V_3$ as an example, writing $V_{mn,m'n'}=\langle m,n|e^{i q_a R_{12}^a}e^{i q_a' R_{12,3}^a}|m',n'\rangle$, we have
\begin{widetext}
\begin{eqnarray}\label{v3}
V_3=\int d^2q_1d^2q_2V_{\vec q_1\vec q_2}\left(\frac{1}{4}V_{30,30}-\frac{\sqrt 3}{4}\left(V_{30,12}+V_{30,12}^*\right)+\frac{3}{4}V_{12,12}\right)
\end{eqnarray}
\end{widetext}
By replacing $V_{\vec q_1\vec q_2}$ with the derived three-body interactions in Eq.(\ref{lll3}) and Eq.(\ref{1ll3}), individual three-body PPs in LLL and 1LL can be computed and compared with those obtained in the literature, which we list in Table (\ref{t2}). They agree perfectly with the most updated values in the literature.
\begin{widetext}
\begin{table*}[ht]
\centering
\begin{tabular}{| c | c | c | c | c | c | c |}
\hline
&&&&&&\\
$\alpha$ & 3 & 5 & 6 & 7 & 8 & 9\\
&&&&&&\\
\hline
&&&&&&\\
$\left(k,l\right)$&$\left(0,1\right)$&$\left(1,1\right)$&$\left(0,2\right)$&$\left(2,1\right)$&$\left(1,2\right)$& $\left(3,0\right)$ $\left(3,1\right)$\\
&&&&&&\\
\hline
&&&&&&\\
$V^{\text{LLL}}_\alpha$ & -0.018101 & 0.003264 & -0.010681 & 0.005945 & -0.004773 &$\left(\begin{array}{cc}
-0.004851 & -0.0007\\
-0.0007 & 0.005224\end{array}
\right)$\\
&&&&&&\\
\hline
&&&&&&\\
$V^{\text{1LL}}_\alpha$ & -0.014684 & -0.005352 & -0.009899 & 0.000451 & -0.000892 &$\left(\begin{array}{cc}
-0.008818 & 0.0007\\
0.0007 & 0.003314\end{array}
\right)$\\
&&&&&&\\
\hline
\end{tabular}
\caption{Values of individual three-body PPs, after stripping away an overall factor from permutation of indices.}
\label{t2}
\end{table*}
\end{widetext}

\section{S3. (Lack of) Galilean invariance in real materials}\label{galilean}
By definition, for a Galilean invariant system its classical Lagrangian $\mathcal L$ satisfies Galilean transformation, which in particular implies that the equations of motion has to be invariant when we move to a different inertial frame with $\vec{\dot x}\rightarrow\vec{\dot x}+\vec v$, where $\vec v$ is the constant velocity difference between the two inertial frames, and $\vec{\dot x}$ is the particle velocity. It is easy to check that the invariance of the equations of motion can only be guaranteed if the kinetic energy in $\mathcal L$ is quadratic in $\vec{\dot x}$ (or that the corresponding Hamiltonian is quadratic in momenta), otherwise the equations of motion will explicitly depend on $\vec v$. For free particles in the limit that the speed of light goes to infinity, the kinetic energy is quadratic in momenta (thus the dispersion relation is a parabola), satisfying the condition for Galilean invariance.

Given a general band dispersion $\epsilon\left(\vec k\right)$ computed from the lattice structure of semiconductors, we can always expand around the bottom or top of the band. Without loss of generality we look at the valence band, for the case that the Fermi level lies above and near the bottom of the band. The band has Galilean invariance if and only if $\epsilon\left(\vec k\right)\sim |\vec{k^*}|^2$, where $\vec{k^*}$ is measured from the bottom of the band. In general, however, terms of higher orders in $\vec {k^*}$ always exist in realistic materials, and Galilean invariance is commonly broken. While such effects can be ignored in most cases and we can still characterize the band dispersion with a single effective mass tensor (given by the inverse of the curvature tensor at the bottom or top of the band), they can still be significant when the Fermi level moves away from the bottom/top of the valence/conduction band. In the case when external magnetic field is applied, the strength of the field also affects the sensitivity to the breaking of Galilean invariance.  
\begin{figure}[htb]
\includegraphics[width=6cm]{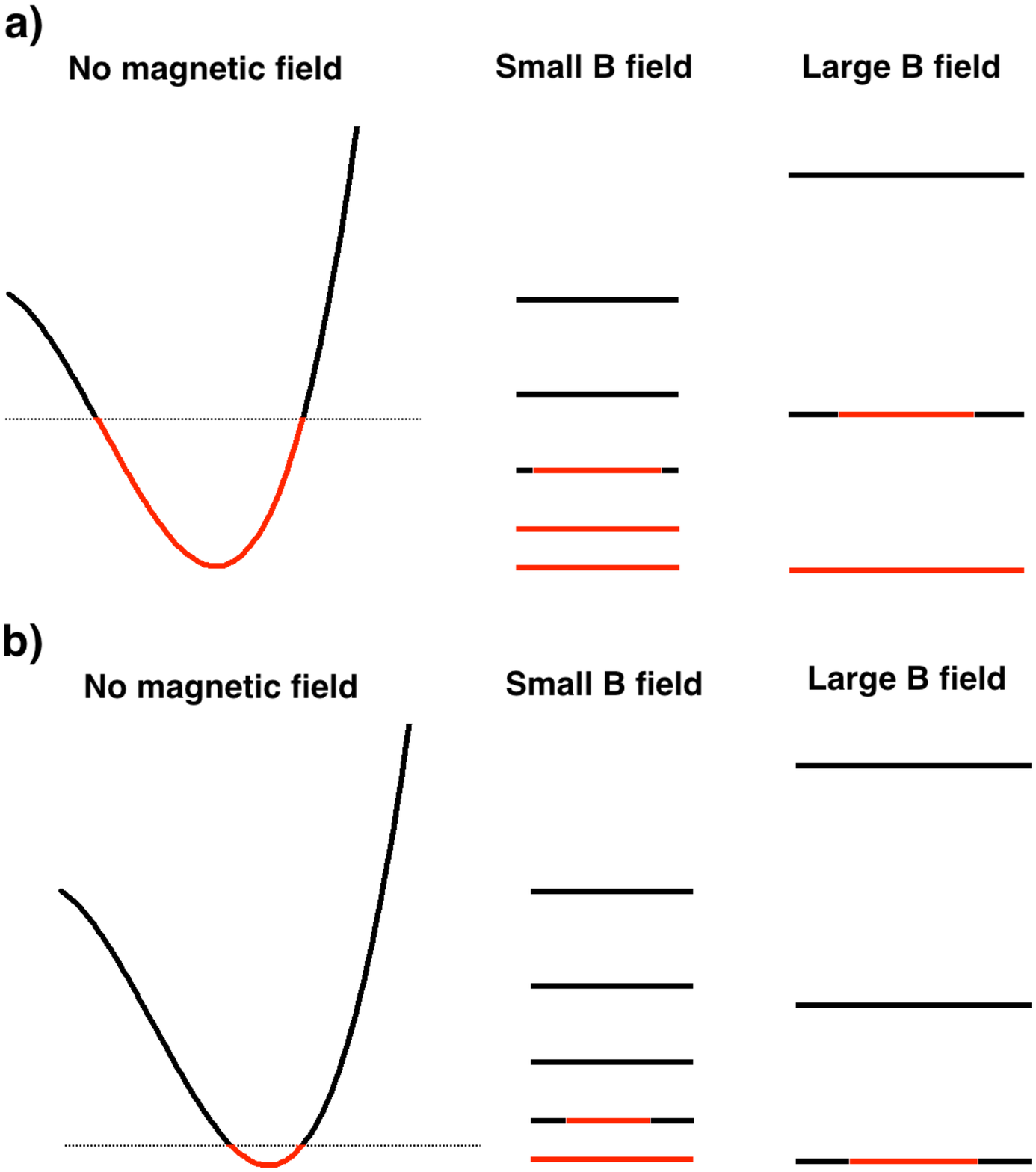}
\caption{Schematic drawing (not to scale) of a non-Galilean invariant band system before and after the application of external magnetic field. The dotted line is the Fermi surface, and the red lines are occupied states.}
\label{s1}
\end{figure}

As one can see from Fig.(\ref{s1}), the external magnetic field completely reorganizes the original band structure, but the LL spacing is inherited from the band dispersion. When the Fermi level is away from the bottom of the band, if the magnetic field is not too strong, electrons in partially filled LL will have significant access to other neighbouring LLs that are not equally spaced. The energy difference between neighbouring LLs increases with stronger magnetic field, though LL mixing is also suppressed. If the Fermi level is close to the band bottom, Galilean invariance is always a good approximation. The detailed quantitative effects of LL energy differences on the effective three-body interactions are presented in the main text.


\begin{thebibliography}{99}
\bibitem{footnote2} The summations in Eq.(\ref{form0}) and Eq.(\ref{form1}) only include particles, thus for graphene the summation needs to be extended to hole LLs with negative energies, which we will not go into details in this work.
\end{thebibliography}
\end{document}